\documentclass[conference]{IEEEtran}
\usepackage{graphicx} \usepackage{amsmath} \usepackage{amssymb}
\usepackage{amssymb}
\usepackage{amsthm} 
\usepackage{algorithm} 
\usepackage{algpseudocode} 
\usepackage{cite} 
\usepackage[caption=false,font=footnotesize]{subfig} 
\usepackage{stfloats}
\usepackage{url}

\usepackage[utf8]{inputenc} 

      \theoremstyle{definition}      \theoremstyle{remark}      \theoremstyle{plain}     

           \newcommand{\mbf}[1]{\mathbf{#1}}          

 \newcommand{\NS}{N_{p}}
 \newcommand{\EP}{E_p}
 \newcommand{\F}[1]{f_{\mbox{\tiny #1}}}
 \newcommand{\NFR}{N_{s}}

 \newcommand{\PD}{P_{\mbox{\scriptsize D}}}
 \newcommand{\PE}{P_{\mbox{\scriptsize e}}}
 \newcommand{\PFA}{P_{\mbox{\scriptsize FA}}}

 \newcommand{\TS}{T_{\mbox{\scriptsize s}}}
 \newcommand{\PDMF}{P_{\mbox{\tiny D}}^{\mbox{\tiny MF}}}
 \newcommand{\PFMF}{P_{\mbox{\tiny FA}}^{\mbox{\tiny MF}}}
 \newcommand{\PDED}{P_{\mbox{\tiny D}}^{\mbox{\tiny ED}}}
 \newcommand{\PFED}{P_{\mbox{\tiny FA}}^{\mbox{\tiny ED}}}
 \newcommand{\PDAD}{P_{\mbox{\tiny D}}^{\mbox{\tiny AD}}}
 \newcommand{\PFAD}{P_{\mbox{\tiny FA}}^{\mbox{\tiny AD}}}
 \newcommand{\PDi}{P_{\mbox{\tiny D}}^{\mbox{\scriptsize i}}}
 \newcommand{\PFi}{P_{\mbox{\tiny FA}}^{\mbox{\scriptsize i}}}

 \newcommand{\NPAD}{N_{p}^{\mbox{\tiny AD}}}
 \newcommand{\NPMF}{N_{p}^{\mbox{\tiny MF}}}
 \newcommand{\NPED}{N_{p}^{\mbox{\tiny ED}}}
 \newcommand{\NP}[1]{N_{p}^{#1}}
 \newcommand{\FK}{f_{k}}
 \newcommand{\TK}{T_{k}}
 \newcommand{\NT}{N_{\mbox{\tiny T}}}

 \newcommand{\mb}[1]{\mbox{#1}}

\begin{document}

\title{IR-UWB Detection and Fusion Strategies using Multiple Detector Types}
\date{\today}
\author{Vijaya~Yajnanarayana,   Satyam~Dwivedi, Peter~H\"{a}ndel\\
  ACCESS Linnaeus Center,\\
  Department of Signal Processing,\\KTH Royal Institute of Technology,   Stockholm, Sweden.\\
  email:\{vpy,dwivedi,ph\}@kth.se}
\maketitle
\begin{abstract}
Optimal detection of ultra wideband (UWB) pulses in a UWB transceiver employing multiple detector types is proposed and analyzed in this paper. To enable the transceiver to be used for multiple applications, the designers have different types of detectors such as energy detector, amplitude detector, etc., built in to a single transceiver architecture. We propose several fusion techniques for fusing decisions made by individual IR-UWB detectors. In order to get early insight into theoretical achievable performance of these fusion techniques, we assess the performance of these fusion techniques for commonly used detector types like matched filter, energy detector and amplitude detector under Gaussian assumption. These are valid for ultra short distance communication and in UWB systems operating in millimeter wave (mmwave) band with high directivity gain. In this paper, we utilize the performance equations of different detectors, to device distinct fusion algorithms. We show that the performance can be improved approximately by $4\,\mb{dB}$ in terms of signal to noise ratio (SNR) for high probability of detection of a UWB signal ($>95\%$), by fusing decisions from multiple detector types compared to a standalone energy detector,  in a practical scenario. 


\end{abstract}




\section{Introduction}
\label{sec:intro}

An ultra wideband (UWB) communication system is based on spreading a low power signal into wideband.  Impulse radio based UWB (IR-UWB) schemes are most popular as they provide better performance and complexity trade-offs compared to other UWB schemes \cite{Win1998}. 

IR-UWB schemes employ narrow impulse signals, which can yield high time resolution, and hence can be used for accurate position localization and ranging. Narrow pulse duration coupled with low amplitude due to the restriction from regulatory agencies like Federal Communications Commission (FCC) makes the detection of these pulses challenging \cite{fcc2002,VJ-3}. Generally, transmit signaling employs multiple pulses and the receiver aggregates certain characteristics from these pulses like energy, amplitude, position, etc., to make statistical inferences on the transmitted information like range (localization) or transmitted symbol value (communication) etc. The performance of the receiver depends on how well the received pulse statistics are utilized for a chosen application.    

\begin{figure*}[t]
\centering
\includegraphics[scale=0.55]{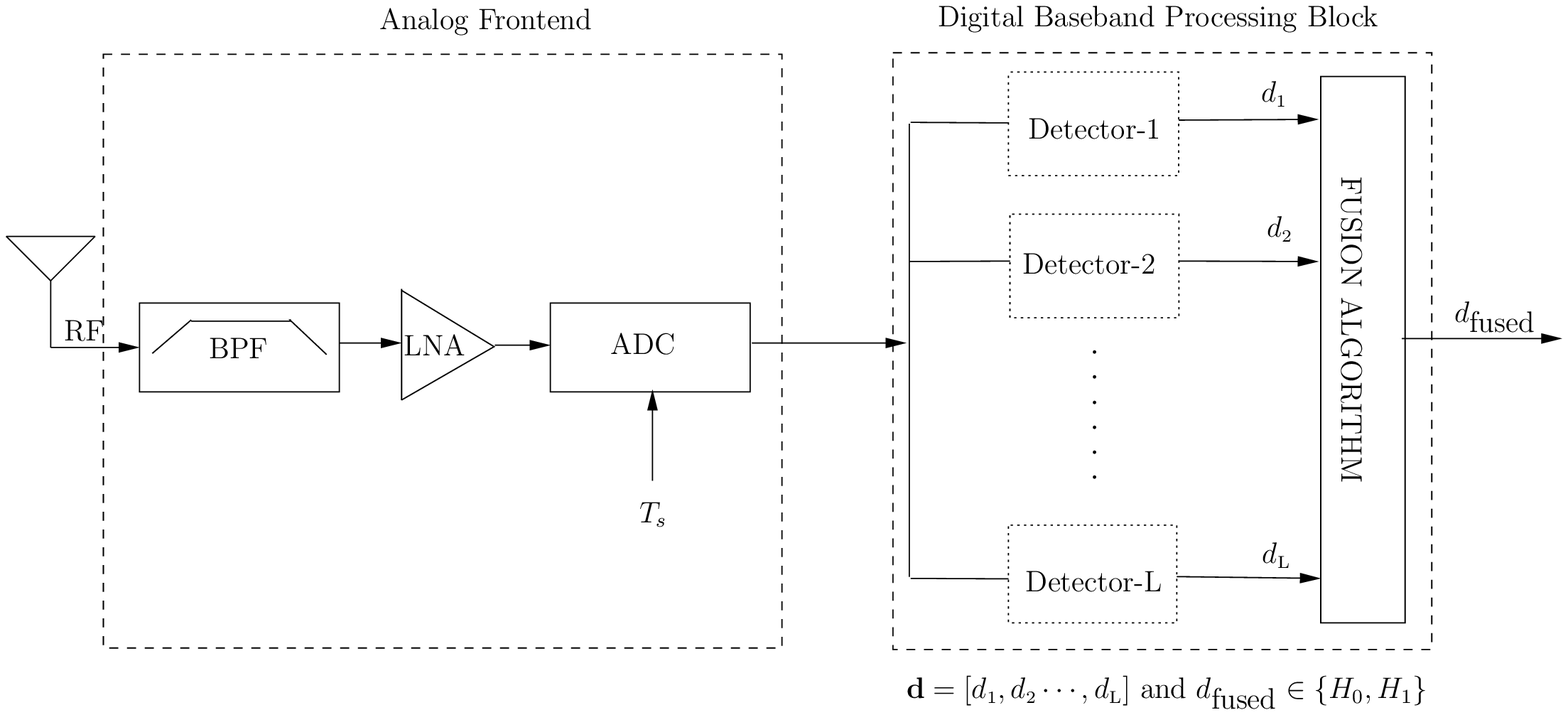}
\caption{Depiction of direct sampling receiver architecture with multi detector fusion. The ($\mbox{Detector-1},\ldots,\mbox{Detector-L}$), are the different detector types available in the transceiver. The $d_i,i\in[1,\ldots,L]$, indicates the binary decisions made by the different detectors\textsuperscript{\tiny[\ref{fnt:1}]} with regard to hypothesis. The $d_{\mbox{\tiny fused}}$, indicates the fused binary decision for the chosen hypothesis.}
  \label{fig:receiver_structure}
\end{figure*}

In this paper, we will consider the structure of a digital sampling receiver shown in Fig~\ref{fig:receiver_structure}. The received signal is filtered by an RF band-pass filter (BPF) and is amplified using a wideband LNA. The signal is then converted into the digital domain by a high sampling rate ADC and digitally processed. IR-UWB pulses are extremely narrow (order of few nano-seconds) and occupy very high bandwidth, therefore high speed ADCs are needed for faithful digital representation of the IR-UWB pulses. The recent progress in the ADC technology, as suggested by  \cite{murmann-adc-survey}, indicates that such high speed ADC having good resolution with signal to noise and distortion ratio (SNDR) of higher than $30~\mb{dB}$  can be achieved for a bandwidth of $10~\mb{GHz}$. This has enabled the digital designs for IR-UWB technology.  In order to exploit the regulatory body specifications optimally, the transceivers must operate at a $3.1-10\,\mb{GHz}$ range or in the unlicensed millimeter wave (mmwave) frequency \cite{fcc2002}.

The digital samples from the ADC will be processed by a digital baseband processing block for detection. In many hardware platforms, a single UWB transceiver mounted on sensors is used for multiple applications like ranging, localization, communication, etc.,  each using particular statistics of the received samples for UWB pulse detection. For example, large distance communication using UWB may employ energy detector over a large number of pulses; whereas short distance tracking application may use amplitude detector on a few pulses. To enable the transceiver to be used for multiple applications, the designers have different types of detectors like amplitude detector, energy detector, etc., built into a single transceiver. Each detector\footnote{Detectors and detector types are interchangeably used. In Fig.~\ref{fig:receiver_structure}, each detector in the set, ($\mbox{Detector-1},\ldots,\mbox{Detector-L}$) are of different type. \label{fnt:1}} uses its own detection algorithm on the received samples to infer a hypothesis from the received samples and report it to the higher layers for further processing. These algorithms are typically implemented in FPGA for faster processing, and hence, only the computed hard or soft-value decisions are available. In some applications, there are no stringent constraints to bind the usage of a particular detector type; for example, demodulation of short range low rate communication data. In these situations, instead of resorting to a single detector type to arrive at the  hypothesis, decision information from all of the different types of detectors can be concurrently utilized to make more informed decision on the hypothesis. This will utilize transceiver infrastructure better, and since every detector decision is new information about the signaled hypothesis, it should yield better reliability and improved performance. 

The proposed transceiver structure shown in Fig.~\ref{fig:receiver_structure}, is applicable to the future evolution of our in-house flexible UWB hardware platform \cite{Ales-2,VJ-3}. This platform can be used for joint ranging and communication applications. The platform has a digital processing section comprising of an FPGA, where the proposed techniques of this paper can be implemented. Even though the applicability of the techniques are demonstrated in simulation, the results provide an early insight in to achievable performance. The variant of the proposed structure in Fig.~\ref{fig:receiver_structure} for hypotheses testing are also employed in \cite{low2003} and \cite{wang-fusion}. In \cite{low2003}, the authors discuss the UWB hypothesis testing for a bank of similar analog detectors, where as in \cite{wang-fusion}, authors proposes a distributed fusion of results from multiple UWB sensors, by allocating the different number of pulses to each sensor, under the constraint of maximum number of allocated pulses, such that the error is minimized. Thus, both are different from the proposed application of this paper. 

In this paper, we formulate a binary hypothesis problem of IR-UWB pulse detection, where decisions from different types of detectors are fused using different fusion methods before deciding on the hypothesis as shown in Fig.~\ref{fig:receiver_structure}. We demonstrate the methods using  three  commonly employed UWB detector-types ($L=3$ in Fig.~\ref{fig:receiver_structure}), having energy detector (ED), matched filter (MF), and amplitude detector (AD) for Detector-1, Detector-2 and Detector-3 respectively. The binary decisions signaling the hypothesis from these three detectors  $\mbf{d}=\left[d_1,d_2,d_3\right]$ are fed to the fusion algorithm to arrive at the binary decision regarding the hypothesis, $d_{\mbox{\tiny fused}}$.

The rest of the paper is organized as follows. In Section \ref{sec:sm}, we will discuss the system model. Here, we will define the signal model which will be used in the rest of the paper. Section \ref{sec:fr}, discusses different fusion strategies. In Section \ref{sec:dp}, we will discuss the analytical expression for  $\PD$ as a function of $\PFA$,  and $\mb{SNR}$ for matched filter, energy detector and amplitude detector for multi-pulse IR-UWB signal. In Section \ref{sec:PE}, we will evaluate the performance of the different fusion strategies. Finally in Section \ref{sec:Conclusion}, we will discuss the conclusions.

\section{System Model}
\label{sec:sm}
We consider a binary hypothesis for detection, with  $H_0$ representing signal is absent and $H_1$ representing signal is present.  Each of the different types of detectors like MF, ED, etc., in the UWB transceiver constructs a test statistic from the received samples, based on which inference is made about $H_0$ or $H_1$ by comparing the test statistic to some threshold, $\gamma$. Different detector types have different ways to construct the test statistic, and thus have varying degrees of performance like probability of detection, $\PD$, probability of error, $\PE$, etc. Apart from the chosen test statistic, the performance of the particular detector also depends on all or few of the parameters listed in the Table~\ref{tab:params}. 
\begin{table}[t]
  \centering
  \caption{Parameters on which detectors performance depends.}
  \begin{tabular}{|l|l|}
    \hline 
    Parameter & Description\\
    \hline 
    $\PFA$ & Probability of false alarm\\
    SNR &  Signal to noise ratio\\
    $\NS$ & Number of UWB pulses used in detection \\
    $\EP$ &  Energy of the UWB pulses\\
    $s(t)$ & Shape of the UWB pulses\\
    \hline
  \end{tabular}
  \label{tab:params}
\end{table}

The transmitted signal under hypothesis $H_1$ consists of $\NT$ frames, such that 
\begin{equation*}
\NT\ge \NP{i} \mb{ }\forall \mb{i}\in [1,2,\ldots,L],
\end{equation*}
where, $\NP{i}$, denotes the number of frames used by Detector-$\mb{i}$ in the hypothesis test. Each frame consists of one IR-UWB pulse, and during hypothesis $H_0$ nothing is transmitted ($\NT$ 
empty frames). Each UWB pulse is of fixed duration, $T$, represented by $s(t)$, sampled at the rate, $1/\TS$, and has $\NFR=T/\TS$, samples. Thus, both hypotheses can be mathematically expressed as 
\begin{equation}
  \label{eq:txsignal}
  \begin{array}{l l}
      \sum\limits_{n=0}^{\NT-1}\sum\limits_{i=0}^{\NFR-1}s(t-nT)\delta(t-nT-i\TS)& \mbox{under } H_1 \\
0 & \mbox{under } H_0     
  \end{array},
\end{equation}
where, $\delta(t)$, denotes the Dirac delta function and the model uses $\NT$ identical frames in each hypothesis test cycle. This is similar to time hopped impulse radio (TH-IR) UWB models proposed in \cite{Win1998,moe_sch_2}, except that we are not considering time hopping, as it has no effect on the statistics collected by the detector across multiple frames. The function,  $s(t-nT)\delta(t-nT-i\TS)$,  represents $i$-th discrete sample of the $n$-th frame under hypothesis $H_1$ and is denoted by $s(n,i)$. The received signal is corrupted by  Gaussian noise. Thus, the received signal used in the hypothesis test under both hypotheses is given by
\begin{equation}
  \label{eq:rxsignal} 
  \begin{array}{l l}
      \sum\limits_{n=0}^{\NT-1}\sum\limits_{i=0}^{\NFR-1}x(t-nT)\delta(t-nT-i\TS)& \mbox{under } H_1 \\
\sum\limits_{n=0}^{\NT-1}\sum\limits_{i=0}^{\NFR-1}w(t-nT)\delta(t-nT-i\TS)& \mbox{under } H_0     
  \end{array},
\end{equation}
where, $x(t)$, is the received pulse shape. The function, $x(t-nT)\delta(t-nT-i\TS)$, represents the $i$-th sample of the $n$-th received frame under hypothesis $H_1$ and is denoted by $x(n,i)$. Similarly, $w(t-nT)\delta(t-nT-i\TS)$, represents the Gaussian noise corresponding to the $i$-th sample of the $n$-th received frame and is denoted by $w(n,i)$. We assume a single-path line of sight (LOS) channel, thus, the received samples, $x(n,i)=\beta s(n,i)+w(n,i)$, where, $\beta$, indicates the path loss.

Typically, the UWB channels are subject to multi-path propagation, where a large number of paths can be observed at the receiver. However, if the transceivers are in close proximity with clear line of sight, the detectors here rely on the first arriving path or LOS, this is in contrast to traditional channel measurement and modeling. If the UWB transceiver is operating at millimeter wave frequencies, due to the combined effect of higher directivity gain due to the RF-beamforming and higher absorption characteristics of the channel results in single-path LOS channels for distances less than $100$ meters.  For the transceiver operating in the frequency band  less than $10~\mb{GHz}$, due to higher reflections, refractions and scattering characteristics of the channel, the assumptions of single-path LOS channel is valid only for extremely short distance of order less than $10~\mb{meters}$ \cite{Ales-2,cotton-channel,Molisch,VJ-3}.  These short distance high speed UWB applications include transferjet and wireless USB (wUSB) \cite{transferjet,wusb}. Also, adopting a simple model proposed here will make the discussion mathematically tractable. Without loss of generality, we use $\beta=1$. In the signal model proposed in  \eqref{eq:txsignal} and \eqref{eq:rxsignal}, we assume perfect synchronization, otherwise there will be degradation of the individual detectors (and fused) performance.

In the next Section, we will discuss the fusion strategies for fusing individual detector decisions (refer to Fig.~\ref{fig:receiver_structure}).

\section{Fusion Rules for IR-UWB Signal Detection }
\label{sec:fr}

\begin{figure}[t]
\centering
\subfloat[AND fusion]{\includegraphics[width=9pc]{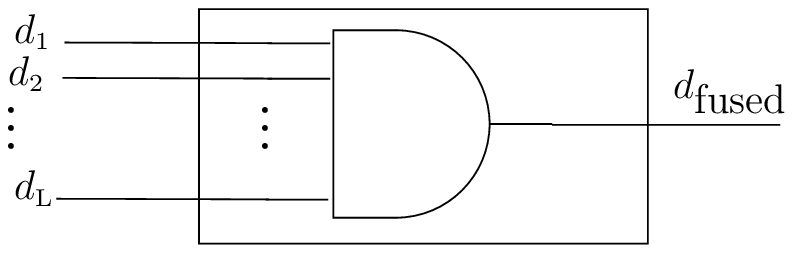}                 \label{fig:AND}}
\subfloat[OR fusion]{\includegraphics[width=9pc]{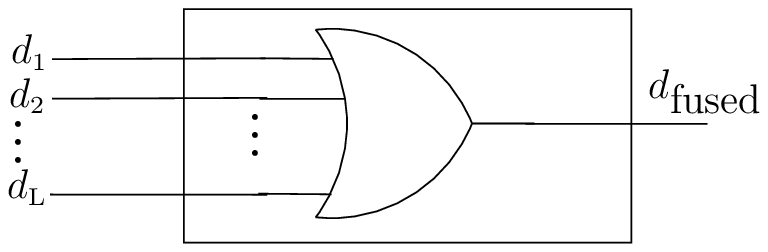} 
  \label{fig:OR}}\\
\subfloat[Fusion based on majority]{\includegraphics[width=9pc]{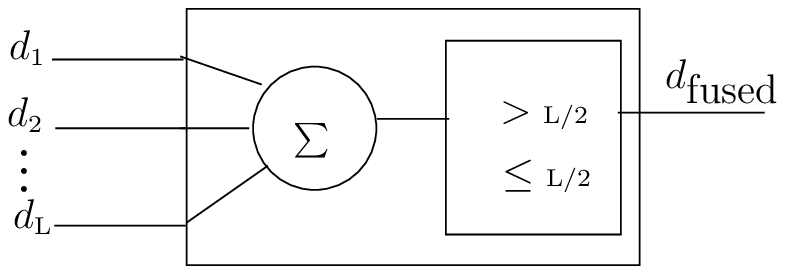} 
  \label{fig:MAJ}}
\subfloat[Maximum a posteriori (MAP) fusion]{\includegraphics[width=9pc]{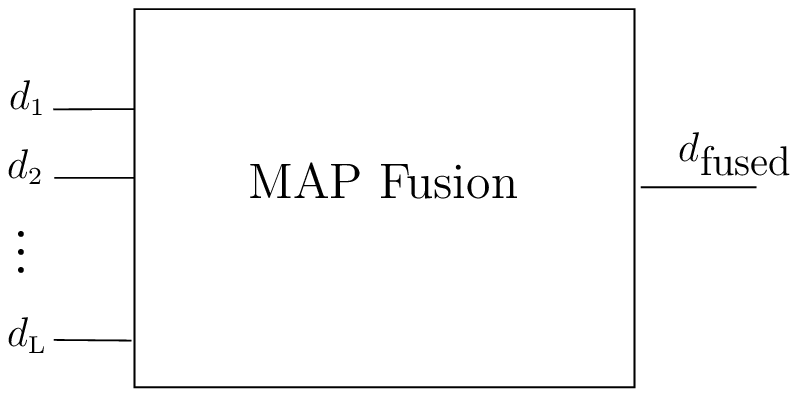} 
  \label{fig:PE}}
\caption{Depiction of different decision fusion methods.}
\label{fig:fusion}
\end{figure}

We consider a general counting rule, that is, deciding for $H_1$, if the sum of the decisions, $\sum_{i=1}^{L}d_i$, exceeds the threshold, $k$.  If we define the decision of the $i$-th detector in the Fig.~\ref{fig:receiver_structure}, as $d_i=0$ and  $d_i=1$ for hypothesis $H_0$ and $H_1$ respectively, then the  special cases of these include simple fusion rules such as ``AND'' ($k=L$), ``OR'' ($k=1$), and ``Majority-Voting'' ($k=L/2$). These fusion rules are depicted in  Fig.~\ref{fig:AND}, Fig.~\ref{fig:OR} and Fig.~\ref{fig:MAJ}. These rules are simple to implement and has been proved to posses robustness features with respect to performance as shown in \cite{Ciuonzo-1,Ciuonzo-2}.

The  counting rule based fusion is biased either toward hypothesis $H_1$ (UWB pulse detection in our model), or  toward $H_0$. For example, fusing using the ``OR'' rule will have superior detection performance, but will also have a larger false alarm rate. Similarly, the ``AND'' fusion rule is  conservative in the UWB pulse detection, but has superior false alarm rate performance. These aspects are further illustrated with numerical examples in the later sections. If we define the mis-classification of the hypothesis as an error and the objective is to minimize the probability of error, $\PE$, then the decision rule discussed above are sub-optimal. This is the motivation for designing a fusion technique that is optimal in probability of error sense. For any prior probability for $H_0$ and $H_1$, the fusion rule that minimizes the probability of error is given by maximum a posteriori (MAP) formulation given below.
\begin{equation}
  \label{eq:intfus0}
  \mb{Pr}(H_1|\mbf{d}) \mathop{\gtrless}_{ H_0}^{H_1} \mb{Pr}(H_0|\mbf{d}).
\end{equation}
Where, $\mbf{d}$, is a $L$-size vector of binary values signaling the hypothesis of the decisions made by different detectors (refer to Fig.~\ref{fig:receiver_structure}). We can write \eqref{eq:intfus0} as
\begin{eqnarray}
  \label{eq:intfus1}
  \log\left(\frac{\mb{Pr}(H_1|\mbf{d})}{\mb{Pr}(H_0|\mbf{d})}\right)\mathop{\gtrless}_{ H_0}^{H_1} 0.
\end{eqnarray}
If we define sets $\mathcal{I}$, $\mathcal{S}_{H_1}$ and $\mathcal{S}_{H_0}$ as
\begin{eqnarray}
  \mathcal{I}&:=&\left\{1,2,\ldots,L\right\}\label{eq:I},\\
  \mathcal{S}_{H_1}&:=&\left\{i:d_i=1\right\}   \label{eq:sh1},\\
  \mathcal{S}_{H_0}&:=&I\setminus\mathcal{S}_{H_1}:=\left\{i:d_i=0\right\}   \label{eq:sh0},
\end{eqnarray}
where, $d_i$, is the binary decision of the detector-$i$ ($i\in\mathcal{I}$), then, 
\begin{eqnarray}
\label{eq:intfus2}
  \mb{Pr}(H_1|\mbf{d})&=&\frac{P_1}{p(\mbf{d})}\mathop{\prod}_{ i\in\mathcal{S}_{H_1}}\PDi\mathop{\prod}_{ i\in\mathcal{S}_{H_0}}(1-\PDi)
\end{eqnarray}
Here, we assumed that the decisions of each of the detectors are independent of each other. $P_1$ is the probability of hypothesis $H_1$ and $p\left(\cdot\right)$ denote the  probability density function (PDF). $\PDi$ is the probability of detection of the $\mb{detector-}i$ in Fig.\ref{fig:receiver_structure}. Similarly, we can write 
\begin{eqnarray}
\label{eq:intfus3}
  \mb{Pr}(H_0|\mbf{d})&=&\frac{P_0}{p(\mbf{d})}\mathop{\prod}_{ i\in\mathcal{S}_{H_1}} \PFi\mathop{\prod}_{ i\in\mathcal{S}_{H_0}}(1-\PFi)
\end{eqnarray}
$P_0$ is the probability of hypothesis $H_0$. $\PFi$ is the false alarm of the $i$-th detector. In many applications such as in communication, hypothesis testing is used for symbol decoding, where both the hypotheses are equally likely. Substituting \eqref{eq:intfus2} and \eqref{eq:intfus3} in \eqref{eq:intfus1} and assuming both hypotheses are equally likely, we get the decision rule as 
\begin{eqnarray}
  \label{eq:intfus4}
    \begin{split}
    \log\left(\frac{\mb{Pr}(H_1|\mbf{d})}{\mb{Pr}(H_0|\mbf{d})}\right)&= \mathop{\sum}_{i\in\mathcal{S}_{H_1}}\log\left(\frac{\PDi}{\PFi}\right)  \\ &\quad+ \mathop{\sum}_{i\in\mathcal{S}_{H_0}}\log\left(\frac{(1-\PDi)}{(1-\PFi)}\right) &\mathop{\gtrless}_{H_0}^{H_1} 0.
\end{split}
\end{eqnarray}

Unlike the counting rule based fusion, the MAP fusion rule employed in \eqref{eq:intfus4}, requires $\PDi$s and $\PFi$s at the fusion center. In practice this is not always available. Also, the fusion rule in \eqref{eq:intfus4}, can be viewed as a weighted counting rule, also known as ``Chair-Varshney'' rule \cite{Varshney-book}. In the next Section, we will derive the detection performance of these detectors, which will be used in the later Sections to evaluate the fusion performance. 

\section{Detector Performance}
\label{sec:dp}

As discussed in Section~\ref{sec:sm}, each transmit frame constitutes a UWB pulse, $s(t)$, sampled at $1/\TS$. We define frame energy, $\EP$ as 
\begin{equation}
  \label{eq:ep}
  \EP=\sum\limits_{i=0}^{\NFR-1}s^{2}(n,i).
\end{equation}
We assume all the frames in the transmission are of same pulse shape, $s(t)$, and energy, $\EP$. As discussed in \eqref{eq:rxsignal}, the received signal under both hypotheses, $H_1$ and $H_0$ is corrupted by AWGN noise samples, $w(n,i)$. We assume that these noise samples are independent and identically distributed (IID) with  $w(n,i)\sim\mathcal{N}\left(0,\sigma^2/\NFR \right)$, where $\mathcal{N}$, denotes the normal distribution, such that the total noise energy in the frame is given by
\begin{equation}
  \label{eq:ner}
  \sum\limits_{i=0}^{\NFR-1}\mbf{E}\left[w^2(n,i)\right]=\sigma^2.
\end{equation}
Here, $\mbf{E}$, denotes the expectation operator. We define signal-to-noise ratio, $\mb{SNR}$, as
\begin{equation}
  \label{eq:snr}
  \mb{SNR}=\frac{\EP}{\sigma^2}.
\end{equation}

\begin{figure}[t]
  \centering
  \includegraphics[width=15pc]{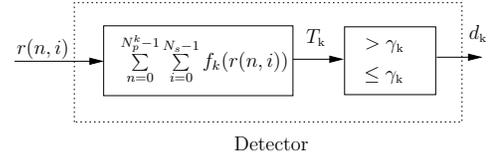}
  \caption{Generic detector structure. The different detector types use different function $\FK(\cdot)$, to construct the test statistic, $\TK$. }
  \label{fig:Detector}
\end{figure}

\begin{figure}[t]
  \centering
  \includegraphics[width=15pc]{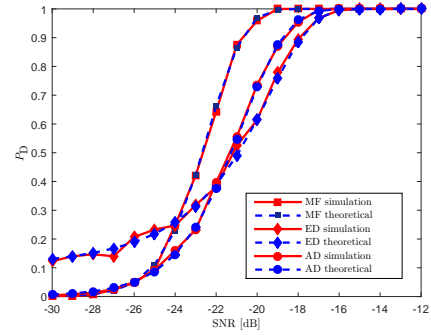}
  \caption{The performance of different detectors in theory and simulation are shown. A normalized second order Gaussian pulse of width $10~\mb{ns}$ sampled at $5~\mb{GHz}$, is used in the simulation. }
  \label{fig:perf_sim_diff_merged}
\end{figure}

Typical detector structure used in Fig~\ref{fig:receiver_structure} is as shown in Fig.~\ref{fig:Detector}. Each detector will construct a test statistic, $\TK$, such that
\begin{equation}
  \label{eq:TK}
\TK=  \sum\limits_{n=0}^{\NP{k}}\sum\limits_{i=0}^{\NFR-1}\FK(r(n,i)),
\end{equation}
and  compare it with a threshold to decide on a hypothesis.  Depending on the test statistic generation function, $\FK(\cdot)$,  we have different types of detectors like matched filter, energy detector, amplitude detector, etc,. In this paper, we use MF, ED, and AD detectors, thus we have, $k\in\{\mb{MF, ED, AD}\}$. $\NP{k}$ denote number of frames used by the detector-$k$, in the hypothesis testing. The $r(n,i)$, denotes the received samples and is equal to $x(n,i)$ and  $w(n,i)$ during hypotheses $H_1$ and $H_0$  respectively.

\subsection{Matched Filter}
\label{subsec:mf}
For matched filter, the test statistic in \eqref{eq:TK} will have
\begin{equation}
\label{eq:Tmf}
  \F{MF}(r(n,i))=r(n,i)s(n,i).
\end{equation}

The performance in terms of probability of detection for matched filter, $\PDMF$, as a function of probability of false alarm, $\PFMF$, and SNR  is shown in Appendix-1 of \cite{VJX-1} to be 
\begin{equation}
  \label{eq:pdmf}
  \PDMF=\mathbf{Q}\left(\mathbf{Q}^{-1} (\PFMF)-\sqrt{\NFR \NPMF \mb{SNR}} \right),
\end{equation}
where $\mathbf{Q}$ is the tail probability of the standard normal distribution.
\subsection{Energy Detector}
\label{subsec:ed}
In energy detector, the test statistic in \eqref{eq:TK} will have
\begin{equation}
  \label{eq:Ted}
  \F{ED}(r(n,i))=r^2(n,i).
\end{equation}
The performance in terms of probability of detection for  energy detector, $\PDED$, as a function of probability of false alarm, $\PFED$, and SNR   is shown in Appendix-2 of \cite{VJX-1} to be 
\begin{equation}
  \label{eq:edint8}
  \PDED=\mathbf{Q}_{\mathcal{X}_{\nu}^2(\lambda)}^{-1}\left(  \sqrt{2\NPED\NFR}\mathbf{Q}^{-1}(\PFED) + \NPED\NFR \right).
\end{equation}
Where $\mathbf{Q}_{\mathcal{X}_{\nu}^2(\lambda)}$ is the tail probability of the non-central chi-square distribution with $\nu=\NPED\NFR$, degrees of freedom, and centrality parameter, $\lambda=\NPED\NFR\mb{SNR}$.

\subsection{Amplitude Detector}
\label{subsec:ad}

In the amplitude detector, the test statistic in \eqref{eq:TK} will have
\begin{equation}
  \label{eq:adint1}
  \F{AD}(r(n,i))=|r(n,i)|.
\end{equation}
The performance in terms of probability of detection for  amplitude detector, $\PDAD$, as a function of probability of false alarm, $\PFAD$,   is shown in Appendix-3 of \cite{VJX-1} to be 
\ifdefined \TWOCOL
\begin{eqnarray}
  \label{eq:adint7}
\begin{split}
 \PDAD&=\mathbf{Q}\left(\mathbf{Q}^{-1}\left({\frac{\PFAD}{2}}\right)-\alpha\sqrt{\NPAD \EP\mb{SNR}}\right)\\
& \quad +  \mathbf{Q}\left(\mathbf{Q}^{-1}\left({\frac{\PFAD}{2}}\right)+\alpha\sqrt{\NPAD \EP\mb{SNR}}\right)\\
\end{split}
\end{eqnarray}
\else
\begin{eqnarray}
  \label{eq:adint7}
\begin{split}
 \PDAD&=\mathbf{Q}\left(\mathbf{Q}^{-1}\left({\frac{\PFAD}{2}}\right)-\alpha\sqrt{\NPAD \EP\mb{SNR}}\right)\\
& \quad +  \mathbf{Q}\left(\mathbf{Q}^{-1}\left({\frac{\PFAD}{2}}\right)+\alpha\sqrt{\NPAD \EP\mb{SNR}}\right)\\
\end{split}
\end{eqnarray}
\fi
where $\alpha$ is defined as 

\begin{equation}
  \label{eq:al}
  \sum\limits_{i=0}^{\NFR-1}s(i)=\alpha \EP.
\end{equation}

As shown by   \eqref{eq:al} and   \eqref{eq:adint7}, the performance of the amplitude detector depends on the shape of the UWB pulse used. We have considered a normalized second order Gaussian pulse as described in \cite{VJ-3}. This is given by
\begin{equation}
  \label{eq:gauss_d2}
  s(t)=-4\pi e^{\frac{-2\pi t^2}{\tau^2}}\left(\frac{-\tau^2+4\pi     t^2}{\tau^4}\right).
\end{equation}

Here $\tau$ can be used to control the impulse spread. Energy normalized pulse, $\EP=1$, with $\tau=3.33~\mb{ns}$, sampled at $5~\mb{GHz}$, will result in $\alpha=4.49$. Thus, for this pulse shape the performance of the amplitude detector is  given by

\begin{eqnarray}
  \label{eq:adint8}
\begin{split}
  \PDAD&=\mathbf{Q}\left(\mathbf{Q}^{-1}\left({\frac{\PFAD}{2}}\right)-4.49\sqrt{\NPAD \EP\mb{SNR}}\right)\\
& \quad + \mathbf{Q}\left(\mathbf{Q}^{-1}\left({\frac{\PFAD}{2}}\right)+4.49\sqrt{\NPAD \EP\mb{SNR}}\right)\\
\end{split}
\end{eqnarray}

From \eqref{eq:pdmf}, \eqref{eq:edint8}, and \eqref{eq:adint8} the performance of matched filter, energy detector and amplitude detector depends on environment (SNR) and on the system configuration or tuning variables like  number of frames considered in the hypothesis testing, $\NS$, and probability of false alarm, $\PFA$. In the matched filter  and energy detector, the performance is agnostic to the system specifications like pulse shape, which are fixed for a given hardware. However, in the amplitude detector, detection performance depends on the shape of the pulse as shown in \eqref{eq:adint7} and \eqref{eq:al}. As discussed in Section \ref{sec:intro}, each detector is pre-configured with detection parameters like $\NS$, $\PFA$, etc., considering a particular application in mind. For an example configuration  shown in Table~\ref{tbl:conf}, the probability of detection, $\PD$, verses SNR using the analytical expression \eqref{eq:pdmf}, \eqref{eq:edint8}, and \eqref{eq:adint8} is as shown in the blue color plots of Fig.~\ref{fig:perf_sim_diff_merged}. 

\subsection{Simulation Study}
\label{sec:ss}
The performance equation for energy detector, \eqref{eq:edint8},   assumes large number of pulses are used in the detection. Similarly, for the amplitude detector, the performance equation, \eqref{eq:adint8}, assumes a UWB pulse shape given in \eqref{eq:gauss_d2}. In this section, we will simulate the detectors and demonstrate the validity of the approximations, for a practical UWB signal setup shown in Table~\ref{tbl:conf}.  We use a  signal model in which each frame is of $10\,\mb{ns}$ duration, having one normalized second order Gaussian pulse as defined in \eqref{eq:gauss_d2} with $\tau=3.33\,\mb{ns}$, sampled at $5\,\mb{GHz}$. The received samples are corrupted by AWGN noise with variance 1/SNR (since pulses are normalized, that is $\EP=1$). Monte-Carlo simulations are done using $1000$ independent realizations. The detector performance in simulations are shown in red,  matches the analytical expressions in \eqref{eq:pdmf}, \eqref{eq:edint8}, and \eqref{eq:adint8}, shown in blue in Fig.~\ref{fig:perf_sim_diff_merged}. Notice that in Fig.~\ref{fig:perf_sim_diff_merged}, different detectors are optimal at different SNR regions. For applications, where there are no stringent constraints to bind the usage to a particular detector type, instead of resorting to a single arbitrary detector to arrive at the hypothesis, decision information from all the detectors can be fused to make more informed decision on the hypothesis. This will utilize transceiver infrastructure better to provide improved performance. We will evaluate the performance of proposed fusion methods  in the next Section.

\begin{table}
  \caption{Configuration of parameters for different detectors used in the fusion.}
  \centering
  \begin{tabular}{|l|l|l|}
    \hline
    Detector Type & $\PFA$ & $\NS$ \\
    \hline
    Matched Filter & $10^{-7}$ & $100$   \\
    Energy Detector & $10^{-1}$ & $1000$   \\
    Amplitude Detector & $10^{-4}$ & $100$   \\
    \hline
  \end{tabular}
\label{tbl:conf}   
\end{table}
 
\section{Performance Evaluation of Fusion Methods}
\label{sec:PE}
\begin{figure}[t]
\centering
\includegraphics[width=20pc]{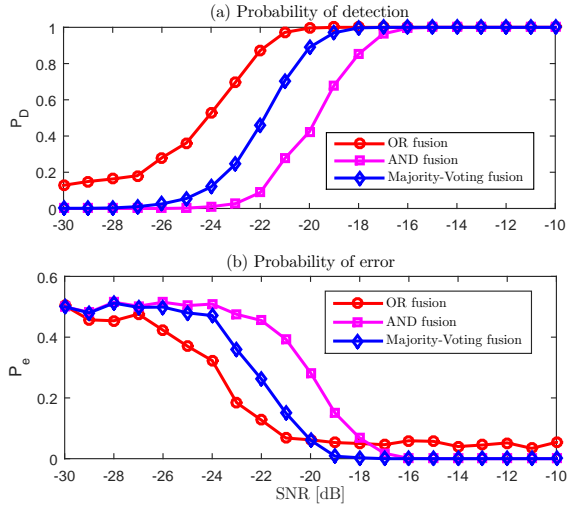}
\caption{Probability of detection and error performance for various fusion techniques using the configuration defined in Table~\ref{tbl:conf}.}
\label{fig:Fig5_c}
\end{figure}

 For counting rule based fusion discussed in Section \ref{sec:fr}, with $L=3$, having MF, ED and AD as detectors  for ``AND'', ``OR'' and ``Majority-Voting'' fusion, we should have $k=3$, $k=1$ and $k>2$ respectively. We performed Monte-Carlo simulations with similar signal configurations described in Section \ref{sec:ss}. We generated $1000$ random signals corresponding to hypotheses, $H_1$ and $H_0$ as defined in \eqref{eq:txsignal}. The probability of correct detection of hypothesis, $H_1$, when $H_1$ was indeed signaled, $\PD$, and the probability of mis-classification of hypotheses, $\PE$, was evaluated using the fusion rules discussed in Section~\ref{sec:fr}. The false alarm, $\PFA$, and number of frames employed, $\NS$, for each detector type are taken from Table~\ref{tbl:conf}. Results for fused probability of detection, $\PD$, and probability of error, $\PE$, are as shown in Fig.~\ref{fig:Fig5_c}a and Fig.~\ref{fig:Fig5_c}b respectively. Notice that for a fixed SNR, the probability of detection is high for the ``OR'' fusion, however, the probability of error is also high due to the higher false alarm rate.

\begin{figure}
\centering
\includegraphics[width=20pc]{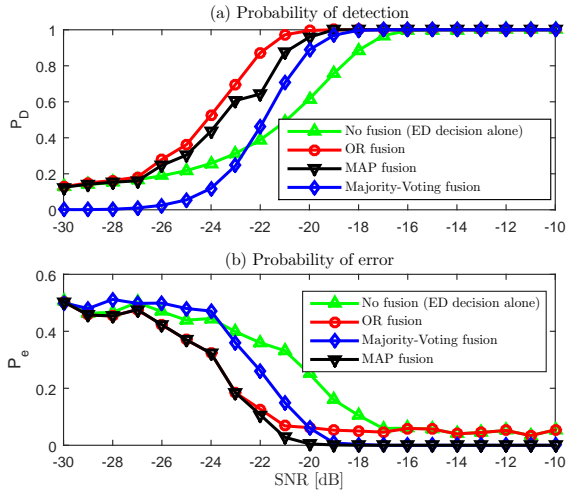}
\caption{Probability of detection and error performance using the fusion rule defined in \eqref{eq:intfus4} (MAP fusion), OR Fusion and ED decision alone without any fusion using the configuration defined in Table~\ref{tbl:conf}.}
\label{fig:Fig6_c}
\end{figure}

The performance is also evaluated using the MAP fusion rule \eqref{eq:intfus4}, for a detector set, (MF, ED, AD), yielding  decision vector, $\mbf{d}$ (refer to Fig.~\ref{fig:receiver_structure}),  with the configuration taken from Table~\ref{tbl:conf}. The probability of detection and probability of error are as shown in Fig.~\ref{fig:Fig6_c}a and Fig.~\ref{fig:Fig6_c}b, respectively. Notice that the MAP fusion method (defined by \eqref{eq:intfus4}) is close to ``OR'' fusion in detection performance, with superior probability of error performance as shown in Fig.~\ref{fig:Fig6_c}b. Comparing the performance of energy detector alone with the MAP fusion rule for multiple detectors in Fig.~\ref{fig:Fig6_c}a and Fig.~\ref{fig:Fig6_c}b, indicates that  a gain of $4\,\mb{dB}$ in terms of signal to noise ratio (SNR) can be achieved for probability of detection greater than $95\%$ with low probability of error ($<5\%$).


\section{Conclusion}
\label{sec:Conclusion}
In this paper, we analyzed the UWB detection performance of matched filter, energy detector and amplitude detector.  We utilized the  performance equations of individual detectors \eqref{eq:pdmf}, \eqref{eq:edint8} and \eqref{eq:adint7} for analyzing the fusion performance. We analyzed the performance in terms of detection probability and probability of error for different fusion methods like  ``AND'', ``OR'' and ``Majority-Voting''. This is shown in Fig.~\ref{fig:Fig5_c}a  and Fig.~\ref{fig:Fig5_c}b. Using the Bayes rule, we derived an optimal fusion rule \eqref{eq:intfus4} for UWB detection, which is optimal in the probability of error sense and compared its performance. This is shown in Fig.~\ref{fig:Fig6_c}a and Fig.~\ref{fig:Fig6_c}b. 

Results indicate that by making a suitable choice of fusion rule, a trade-off between detection and false alarm can be achieved. For example, Fig.~\ref{fig:Fig5_c}a, shows that OR fusion is more biased toward detection, however, it also results in higher errors (due to false alarms, refer to Fig.~\ref{fig:Fig5_c}b). If the error performance is critical for the UWB application, then MAP fusion formulation  gives superior performance in terms of errors as shown in Fig.~\ref{fig:Fig6_c}b. In general, if there are multiple detectors available in the UWB transceiver platform, then decision information from these detectors can be concurrently utilized and intelligently fused based on the application criteria to make a  more informed decision on the hypothesis.

\bibliography{my}
\bibliographystyle{IEEE}

\end{document}